\documentclass[pdflatex,iicol]{sn-jnl}

\usepackage[section]{placeins}
\usepackage{textgreek}
\usepackage[version=4]{mhchem}
\usepackage{siunitx}
\DeclareSIUnit\molar{\textsc{m}}
\usepackage[switch]{lineno}

\jyear{2023}%

\begin{document}

\title[Two dimensional Alkaline Vents]{Alkaline Vents Recreated in Two Dimensions to Study pH Gradients, Precipitation Morphology and Molecule Accumulation}

\author[1]{\fnm{Maximilian} \sur{Weingart}}\email{m.weingart@physik.lmu.de}
\author[2]{\fnm{Siyu} \sur{Chen}}\email{siyu.chen@tum.de}
\author[3]{\fnm{Clara} \sur{Donat}}\email{clara.donat@tum.de}
\author[4]{\fnm{Vanessa} \sur{Helmbrecht}}\email{v.helmbrecht@lrz.uni-muenchen.de}

\author[4,5]{\fnm{William D.} \sur{Orsi}}\email{w.orsi@lrz.uni-muenchen.de}

\author[1]{\fnm{Dieter} \sur{Braun}}\email{dieter.braun@lmu.de}

\author*[3,2]{\fnm{Karen} \sur{Alim}}\email{k.alim@tum.de}

\affil[1]{\orgdiv{Systems Biophysics and Center for NanoScience (CeNS)}, \orgname{Ludwig-Maximilians Universität München}, \orgaddress{\street{Amalienstraße 54}, \city{München}, \postcode{80799}, \state{Bavaria}, \country{Germany}}}

\affil[2]{\orgname{Max Planck Institute for Dynamics and Self-Organization}, \orgaddress{\street{Am Faßberg 17}, \city{Göttingen}, \postcode{37077}, \state{Niedersachsen}, \country{Germany}}}

\affil[3]{\orgdiv{CPA and Department of Bioscience, School of Natural Sciences}, \orgname{Technische Universität München}, \orgaddress{\street{Ernst-Otto-Fischer-Straße 8}, \city{Garching b. München}, \postcode{85748}, \state{Bavaria}, \country{Germany}}}

\affil[4]{\orgdiv{Department of Earth and Environmental Sciences}, \orgname{Ludwig-Maximilians Universität München}, \orgaddress{\street{Richard-Wagner Straße 10}, \city{München}, \postcode{80333}, \state{Bavaria}, \country{Germany}}}

\affil[5]{\orgdiv{GeoBio-CenterLMU}, \orgname{Ludwig-Maximilians Universität München}, \orgaddress{\street{Richard-Wagner Straße 10}, \city{München}, \postcode{80333}, \state{Bavaria}, \country{Germany}}}


\abstract{
Alkaline vents (AV) are hypothesized to have been a setting for the emergence of life, by creating strong gradients across inorganic membranes within chimney structures. In the past, 3-dimensional chimney structures were formed under laboratory conditions, however, no in situ visualisation or testing of the gradients was possible.

We develop  a quasi-2-dimensional microfluidic model of alkaline vents that allows spatio-temporal visualisation of mineral precipitation in low volume experiments. Upon injection of an alkaline fluid into an acidic, iron-rich solution, we observe a diverse set of precipitation morphologies, mainly controlled by flow-rate and ion-concentration. Using microscope imaging and pH dependent dyes, we show that finger-like precipitates can facilitate formation and maintenance of microscale pH gradients and accumulation of dispersed particles in confined geometries.

Our findings establish a model to investigate the potential of gradients across a semi-permeable boundary for early compartmentalisation, accumulation and chemical reactions at the origins of life.
}


\maketitle
\newpage
\section{Introduction}\label{sec1}


Investigating the origins of life necessarily leads to the question which possible locations could have provided promising conditions for the emergence of living systems on a prebiotic Earth. One of the theories today focuses on  alkaline vents (AVs) as a promising setting, providing unique chemical and geo-physical conditions \cite{Russell1997, Martin2008, Sojo2016}.
AVs on the prebiotic seafloor are predicted to have formed chimney structures, where warm alkaline fluids were exhaled into an acidic ocean enriched in dissolved iron(II) \cite{Poulton2011, Nisbet1985}. Upon contact of the fluids highly reactive Fe(Ni)S, minerals precipitate and enclose the alkaline fluid stream to form a tubular network that grows into chimney-like structures \cite{ Preiner2018, Sojo2016}. The precipitate thereby acts as a permeable mineral membrane \cite{Russell2010, Moller2017} that allows for the generation of steep gradients in pH, redox-potential and temperature, thus providing not only the non-equilibria that might have been crucial for the emergence of life \cite{Dirscherl2023, matreux2023formation} but also a potential energy source for early life forms \cite{Moller2017, Martin2008}. The mineral surfaces, at the same time, could serve as a catalytic site for abiotic organic synthesis reactions \cite{Preiner2018, Sleep2004} while the inflowing water would provide a constant source of chemical nourishment \cite{Martin2008}. As shown in recent work \cite{Helmbrecht2022} vent minerals could have also facilitated the localisation and accumulation of dissolved RNA molecules. The physical non-equilibrium conditions provided by temperature gradients between the fluids, can further boost accumulation of dissolved species and increase their local concentration manifold \cite{Duhr.2006, Baaske.2007}. The combination of those characteristics makes AVs also a very interesting location for inorganic catalytic reactions or polymerisation or replication \cite{Dass2022, Dirscherl2023, Mast.2010, Mast2013}. On the other hand, the gradients in pH, separated by a semipermeable membrane, could have worked as an electro-chemical reactor, providing a precursor setting for modern autotrophic cells \cite{Sojo2016, Sojo2019, Martin2008, Russell1997, Russell2010}. Which of the conditions provided by alkaline vents, however, are critical or sufficient to drive molecular evolution at the origins of life is mostly unknown.

    \begin{figure}[t]
		\centering
		\includegraphics[width=\linewidth]{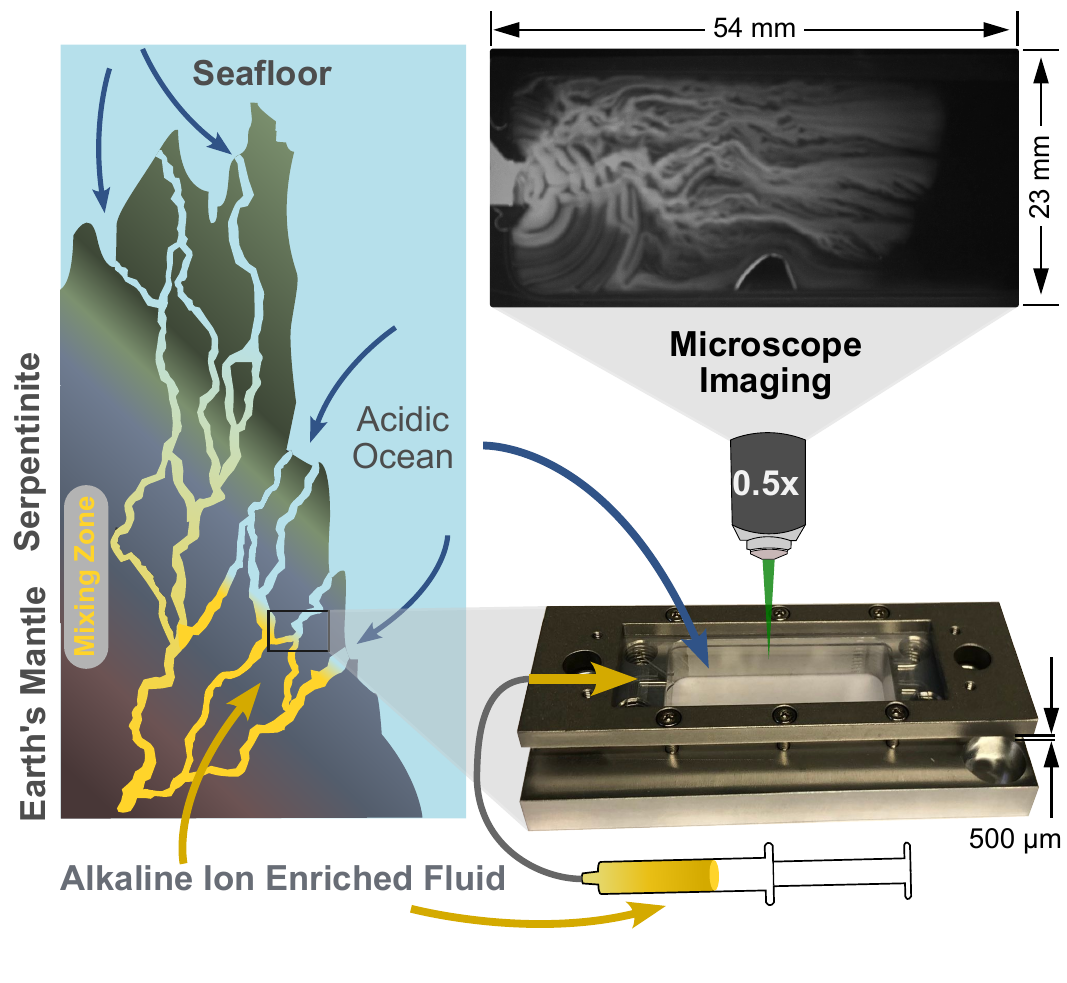}
		\caption{Mixing of alkaline and acidic solutions in seafloor rock pores is mimicked in a microfluidic flow cell. Acidic ocean water percolates the narrow cracks and pores of the seafloor and is converted to an ion enriched and strongly alkaline fluid by interaction with the surrounding rock. When this fluid is exhaled back into the ocean, rapid mineral precipitation happens upon contact with the ocean water. To mimic this scenario in the lab, an alkaline sodium-hydroxide solution is pumped at a controlled flow rate into a flat microfluidic chamber pre-filled with an acidic iron-solution. Microscope imaging with pH dependent dyes allows the visualisation of different morphologies, depending on inflow-rate and OH- concentration, and the assessment of emerging pH gradients.}
		\label{fig_motivation}
	\end{figure}


A well studied example of AVs formed at moderate temperatures is the Lost City hydrothermal field near the mid-Atlantic spreading centers \cite{Kelley2001, Kelley2005}. In general, these vents emerge when ocean water percolates into cracks and fractures of Earth's upper mantle, which typically comprises minerals of ultramafic rock. Upon contact with water a metamorphic chemical reaction (serpentinisation) converts the ultramafic rock to serpentinite \cite{Sleep2004, Bach2006}. 
Even though Lost City can be used as an example for the essential nature of AVs, the geochemical conditions are assumed to have been quite different on a prebiotic Earth: The early ocean likely has been oxygen-free and saturated with \ce{CO2} from the atmosphere \cite{Russell2010}, hence, slightly acidic \cite{Russell1997, Macleod1994, Halevy2017}, and enriched with up to milli-molar concentrations of dissolved \ce{Fe^2+} \cite{Russell2010, Swanner2020, Poulton2011, Russell1997}. Serpentinisation and the overall precipitation mechanism, however, were likely prevalent on an early Earth. And since ultramafic intrusions could have been more frequent \cite{Nisbet1985}, probably even more abundant. NASA undertook several studies to recreate the conditions of prebiotic AVs in the laboratory, but has so far only investigated the structures in three dimensions \cite{Herschy2014, Barge2015b, Burcar2015, Barge2019}.


Controlled precipitation reactions have a long history, reaching back to early alchemists, and are now comprised in the field of chemobrionics \cite{Barge2015}. The resulting structures, often referred to as chemical gardens, can be seen as a simple example of self-organising systems that  exhibit all kinds of different morphologies under different chemical and physical conditions \cite{Barge2015, Cartwright2002}. Apart from precipitation and crystallisation, structure growth is driven by fluid advection through either active injection \cite{Stone2005, Haudin2014, Balog2020, Bere2021} or osmosis and convection \cite{Sainz-Diaz2018, Ding2022}. In the past years, they have also been increasingly studied in quasi-2-dimensional settings using microfluidic devices like Hele-Shaw flow cells \cite{Haudin2014, Schuszter2016, Wang2020, Rocha2022}. The focus is mostly on reaction parameters that allow ordered structures like hollow cylinders, but also on mathematical modeling of the resulting patterns \cite{Stone2005, Comolli2021, Rocha2022}. Because of their apparent similarity to living organisms \cite{Barge2015} there was huge interest in chemical garden structures in the early stages of the origins of life field \cite{Leduc1911, Cartwright2002}, but had been neglected since, until NASA researchers started to recreate hydrothermal vent analogs in the lab \cite{Herschy2014, Barge2015b}. However, so far none of the chemical garden studies in confined geometries has attempted to recreate the conditions of the origins of life on Earth.


In our experiments we use a thin, translucent microfluidic cell built up by a cutout from a Teflon sheet that is sandwiched between two sapphire slides. The rectangular chamber is filled with an acidic iron solution to simulate the ocean and a sodium hydroxide solution injected by a syringe pump simulates the venting of alkaline fluid (see Fig.~\ref{fig_motivation}). This setup of the flow cell allows for microscope imaging in bright-field and dark-field mode. Although the thin mineral membranes formed in hydrothermal vents have been studied in microfluidics before \cite{Moller2017, Sojo2019, Wang2020b, Batista2015, Wang2019}, only controlled single membranes were created by laminar co-flow of the fluids. Our experiments, here, are designed to provide a two dimensional representation of the three dimensional physiology of AVs, where acidic and alkaline fluids come into contact.

In this paper, we show that our setup enables us to visualize the formation of pH gradients in a quasi-2-dimensional model of hydrothermal vents to study the effects of physical parameters like flow-rate and fluid concentration on resulting morphologies and emerging gradients. Our results show that finger-like precipitation patterns at mediate flow and concentration facilitate the formation of pH gradients while at the same time allowing for the accumulation of dispersed particles on the reactive mineral surfaces. The confined microfluidic chamber can therefore not only be seen as a model for the geological analogue, but in fact also as a simulation of potential mixing of the fluids that could already start in the cracks and fractures of the seafloor even before the alkaline fluid is vented into the ocean. Our setup therefore provides a way to simulate chemio-physical hydrothermal vent experiments in the lab in low volume.

\section{Results}\label{sec2}


To mimic mineral precipitation in hydrothermal vents, we used a custom built microfluidic flow-cell to inject alkaline fluids into acidic ocean water (Fig. S1). This, on the one hand, enables us to get a spatio-temporal visualisation of the resulting morphologies, and on the other hand also serves as a model for potential precipitation processes already starting in the narrow rock pores in the sea floor. The microfluidic chamber of \SI{500}{\micro\meter} thickness is pre-filled with an acidic, iron containing ocean analogue, in which the alkaline fluid is injected at a constant flow-rate using a high precision syringe pump. Gradients and morphologies can be visualised with the aid of pH sensitive dyes immersed into both fluids.

    \begin{figure*}[ht!]
		\centering
		\includegraphics[width=\linewidth]{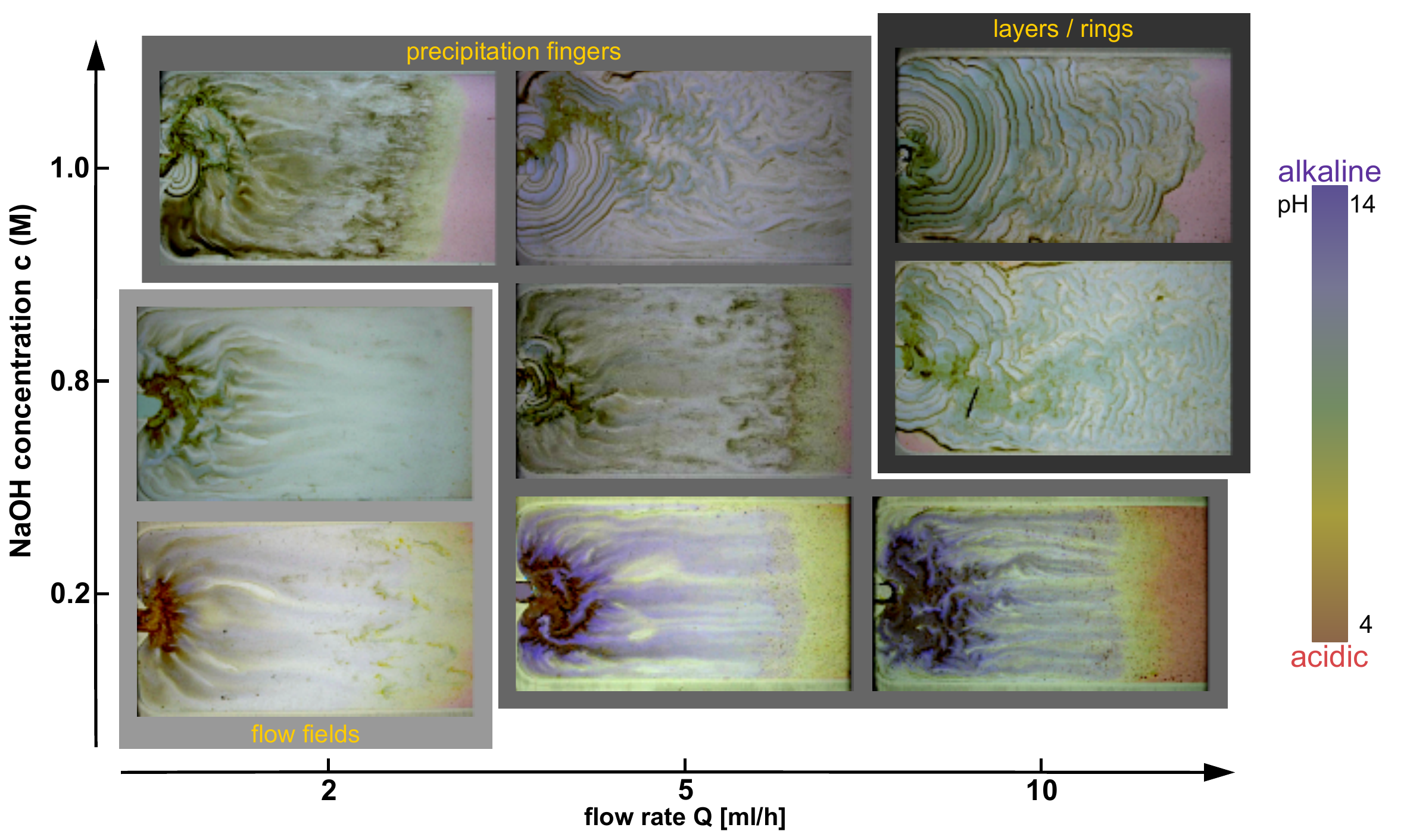}
		\caption{Variation of inflow-rate and concentration of the alkaline NaOH solution yields three dominant morphologies. ``Layers/rings" form at high flow and concentrations, whereas diffusive fields persist at low values. In between both regimes, precipitate fingers form that allow the formation and maintenance of steep pH gradients. At the higher limit, the acidic ocean is almost completely repressed by the precipitation front and at the lower limit, both solution mix by diffusion due to weak precipitation. Therefore, no gradients can be formed or maintained in both limiting morphologies.}\label{fig_flow_conc}
	\end{figure*}


\subsection{Variation of pH and Flow-Rate}
We found that the physical parameters influencing the resulting morphology are mainly flow-rate and the alkalinity (i.e.~the concentration) of the inflowing sodium hydroxide solution. By variation of these two parameters, we defined boundaries for each regime at which the final morphology is not altered by further changes in one of the parameters. 
The influence of flow rates and concentrations on the resulting morphologies is visualised using a universal pH indicator. The steady-state images, in which further inflow does not alter the morphology any more are shown in Fig.~\ref{fig_flow_conc}. Different inflow rates of \SI[parse-numbers = false]{2.0-10.0}{\milli\liter\per\hour} and alkaline concentrations from \SI[parse-numbers = false]{0.2-1.0}{\molar} result in a diverse set of morphologies, from which three dominant structural patterns can be identified: ``layers/rings", ``precipitation fingers" and ``flow fields".
Diffusive fields form as large almost homogeneous areas of both fluids due to weak precipitation at low flow-rate and/or low inflow concentrations, whereas  ``layers and rings" emerge as thick layered precipitates completely excluding the ocean fluid at high flow and strongly alkaline pH. In between the limits of both regimes, we observe ``precipitation fingers" as parallel structures that maintain stable precipitation patterns. Higher concentrations of in-flowing sodium hydroxide, together with high flow rates, result in faster precipitation and stronger precipitate formation. The motion of the fluid flow entering the chamber is purely advection driven. Lowering flow and/or influx concentration further the initial motion gets increasingly diffusion governed. Therefore, the diffusive fields takes the longest time to reach the steady-state, since even after complete filling of the chamber, intermixing of the fluids still continues via diffusion across the weak precipitate structures.

    \begin{figure*}[ht!]
		\centering
		\includegraphics[width=\linewidth]{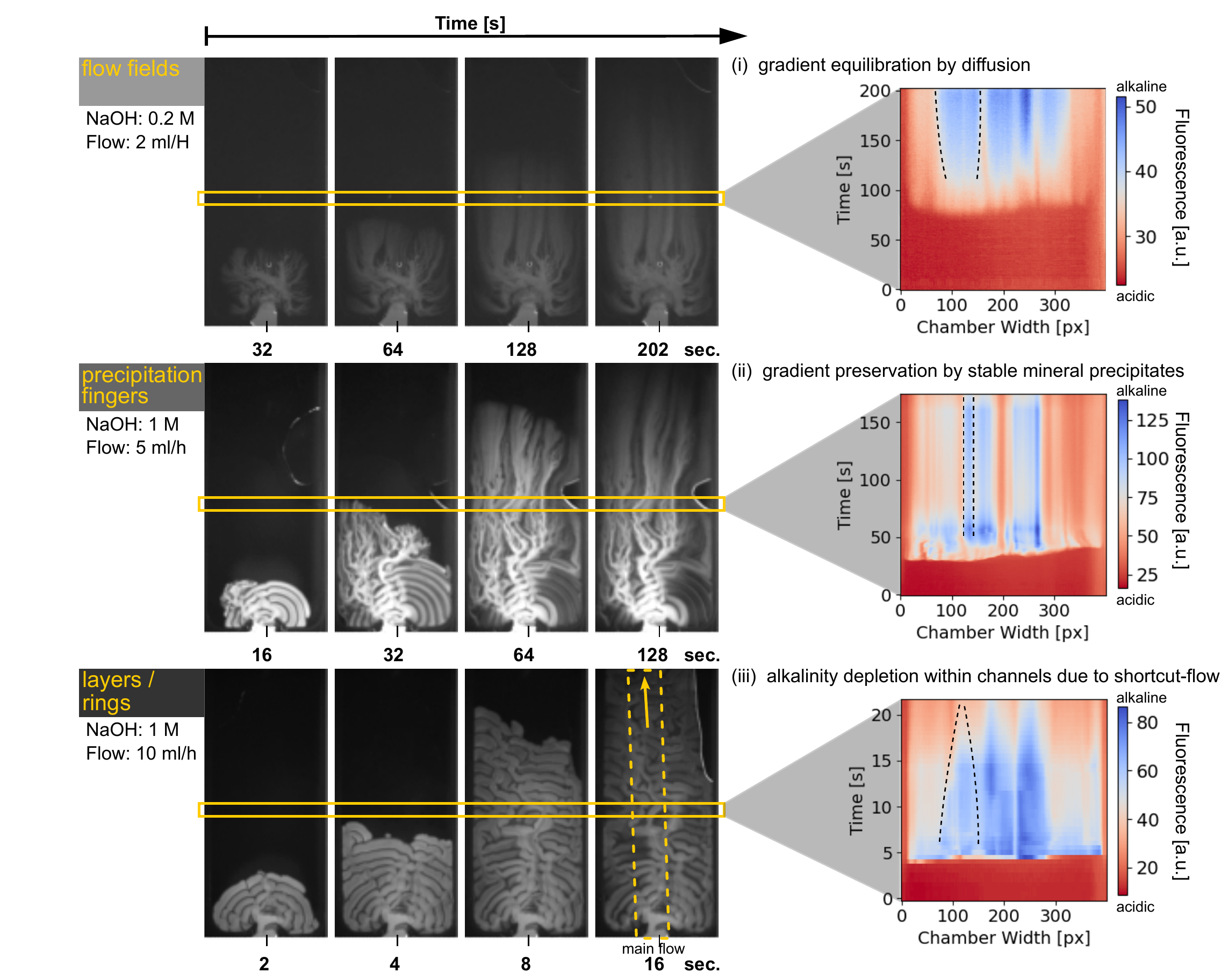}
		\caption{Changes in inflow-rate and concentration correspond to a timescale shift in morphology formation. Each of the three time-series represents conditions that are dominated by one of the three morphologies. ``layers/rings" form predominately within the first \SI{20}{\second}, then converge to the finger structures between roughly 30 and \SI{80}{\second} and finally fade into the flow fields after around \SI{100}{\second}. On the right an average of 20px in height over the cross-section of the chamber is plotted over time: (i) No steady-state is reached and the fluids intermix by diffusion over time (blue areas become wider). (ii) Stable gradients are formed between red (acidic) and blue (alkaline) areas and are maintained over time (straight blue stripes at persistent width). (iii) In steady-state a dominant flow through the layers forms and fluorescence fades in adjacent structures, indicated by narrowing of the blue (alkaline) areas. }\label{fig_time}
	\end{figure*}


\subsection{Time-Dependent Morphology Formation}
Variation of flow-rate and concentration of the injected fluid showed that the three dominant morphologies emerge on different, subsequent time scales. An even clearer visualisation can be seen in Fig.~\ref{fig_time}, where the alkaline areas for each of the three morphologies are highlighted by a pH dependent fluorescent dye (BCECF). Among these three time-dependent patterns, ``layers/rings" form within the first \SI[parse-numbers = false]{10-15}{\second} and fill up the complete chamber. For the mid-regime initially thick layered precipitates grow until around \SI{30}{\second},  at which point the bulk flow is divided into multiple smaller flow trajectories that form the characteristic precipitation fingers. With increasing time these precipitations get weaker and the fingers start to diffuse out perpendicularly to the flow direction until the top end of the chamber is reached. Similarly, in experiments revealing ``flow fields" as dominating morphology, we observe the shaping of weak and thin finger-like precipitates within the first \SI{120}{\second} until the diffusive structure of the homogeneous ``flow fields" dominates, due to mitigating precipitation.
By varying \ce{OH-}-ion concentration and inflow rate, we can control the time-scales for the mineral precipitation pattern formation. The formation time-scale gets longer for each of the dominating patterns and ranges from a few seconds (layers/rings) over \SI[parse-numbers = false]{1-2}{\minute} (precipitation fingers) to above \SI{3}{\minute} (flow fields), where no stable steady-state could be observed within the measurement time. Comparing all three experiments, which exhibit one dominant morphology each, (see Fig.~\ref{fig_time}) we can assess that the overall vent formation also scales over time.


\subsection{pH dependent Dye and Fluid Flow}
Apart from mapping the areas of different pH, the fluorescent dye also allows the tracing of flow and ion interchange over time (see Fig.~\ref{fig_time}). This is essential for the observation of two major features: gradient formation and their subsequent maintenance after entering the morphological steady state. 
Especially in the ``layers/rings" it can be seen that over time a main flow dominates through the center of the thick layers leading to fading intensity of the fluorescent dye in the areas farther outside. This indicates a decrease in pH most likely due to ionic interchanges between the fluid and the mineral precipitates.
However, this effect cannot be observed in the ``precipitation fingers" as the flow is not bypassing earlier formed structures but rather divides into the smaller sub-flows forming the characteristic finger structures. The constant ion-inflow maintains the fluorescence and, hence, the pH gradients even after reaching the steady state.
The ``flow fields" on the other hand, show fading of the fluorescent dye by diffusing into the remaining ocean liquid due to weak and unstable mineral structures. This happens even though a similar sub-flow division as for the ``precipitation fingers" can be seen in the beginning (Fig.~\ref{fig_time}). Because of this ongoing mixing of the two fluids by diffusion, all initially emerging gradients are equilibrating in the ``flow fields" morphology.


\subsection{Potential for Gradient Formation}
The potential of each morphology for the formation of gradients can be assessed from the final vent morphology of each time evolution in Fig.~\ref{fig_time}.
The ``layers/rings" structure do not allow for any gradients to form between alkaline fluid and acidic ocean analogue, since all of the acidic liquid in the chamber is either consumed in precipitation or depleted from the chamber as it fills up with precipitate and alkaline solution. The only possible gradients would be the ones between shortcut-flow and areas of fading pH as described above. However, those gradients would not be maintained and separated by a mineral membrane and thus equilibrate quickly by diffusion. This effect is shown in the plot of a cross-section through the chamber over time plot (Fig.~\ref{fig_time} (iii) right): The width of the blue (alkaline) decreases, as the fluorescence of the outer areas is depleted and a main flow forms.
Same is valid for the ``flow fields" structures, since no steady-state can be reached within the time of the experiment such that the fluids keep intermixing through the weak membranes forming between \SI{30}{\second} and \SI{200}{\second}. Thus, the pH is equilibrating gradually, which can be seen by the faint edges of the precipitates in the dark field images and the widening width of blue/white (alkaline) areas in the cross-section over time plot (Fig.~\ref{fig_time} (i) right).
In contrast to that, the ``precipitation finger" structures show a clear pH difference of the fingers to the acidic solution in between. The strong precipitates act as a semi-permeable mineral membrane that keeps high and low pH areas separated and together with the continued inflow of \ce{OH-} ions maintains the gradient even in the final state of the morphology. This can clearly be seen by the straight blue (alkaline) stripes in the cross-section vs.~time plot (Fig.~\ref{fig_time} (ii) right). Their width does not significantly vary over time, indicating a stable gradient between red (acidic) and blue (alkaline) areas. The maintenance of pH gradients across stable semi-permeable membranes is a crucial feature for life as we know it today. The characteristic of pH gradient formation can only be observed for the ``precipitation fingers" at mediate flow and concentration of inflowing fluid. Modern compartments, like cells, provide a similar setting: an alkaline interior separated by a membrane from an acidic surrounding. To keep the gradients across the membrane up, however, for cells mechanisms like trans-membrane machines are needed to pump protons out of the cell (see Fig.~S2). In contrast, the finely finger-like structures in the microfluidic morphology, would not require any active means or proton pumping as long as the flow of alkaline fluid continues.


    \begin{figure}[ht!]
       \centering
        \includegraphics[width=\linewidth]{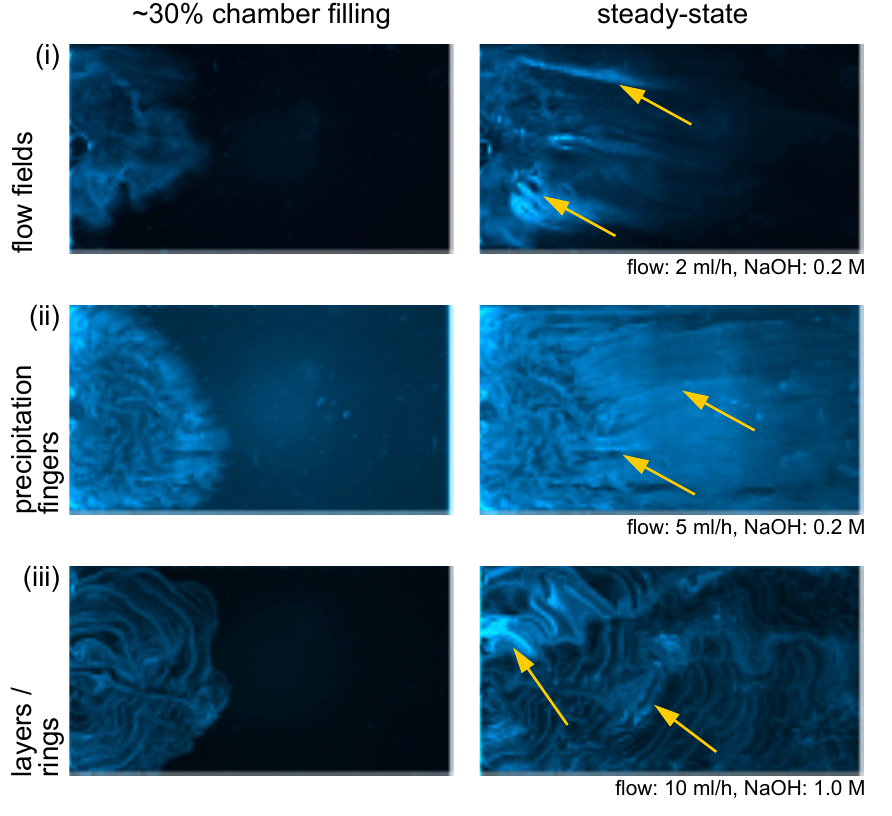}
        \caption{Accumulation of fluorescent beads (\SI{10}{\micro\meter}) in the different dominant morphologies, ``layers/rings" (i), ``precipitation fingers" (ii) and ``flow fields" (iii). Strongest local increase of fluorescence is observed in the ``precipitation fingers" (ii) suggesting the accumulation of dispersed beads.}\label{fig_beads}
    \end{figure}
    
\subsection{Bead Accumulation}
The next step in testing the capabilities for compartmentalisation, is to check for potential accumulation of dissolved species in the fluids. We tested our different morphologies with fluorescent beads of \SI{10}{\micro\meter} diameter. In Fig.~\ref{fig_beads} three experiments are shown with conditions yielding the dominant morphologies, but with dispersed fluorescent beads mixed in the inflowing fluid. It can be seen that beads are getting trapped inside the morphologies upon precipitation. In the ``flow fields" (Fig.~\ref{fig_time} (i)) a very weak fluorescence signal indicates hardly any bead accumulation, which is most likely due to the weak precipitation and pH gradients. Although there appears to be a higher fluorescence signal in the ``layers/rings" (Fig.~\ref{fig_beads} (iii)) due to their high inflow rate, most of the beads fade again over time as they are most likely flushed out. Nevertheless, beads are locally accumulated in some of the grooves along the fluid paths, however, they would not be subjected to any gradients or continued feeding after a shortcut-flow forms after some time (see \ref{sec2}). 
In the ``precipitation fingers" structures (Fig.~\ref{fig_beads} (ii)), we see that some beads are getting trapped in the precipitate upon formation, which forms local clusters of accumulated particles, increasing in fluorescence over time, which suggest continued accumulation over time as the influx persists. In this way, spots of locally increased concentration of accumulated particles form, that persist even after the precipitation process is finished.

\section{Methods}\label{sec11}

The microfluidic setup (Fig.~S1~a) is a custom made flow cell consisting of a two-dimensional, rectangular geometry (\SI{16}{\milli\meter}\,$\times$\,\SI{40}{\milli\meter}) with in-/outlet channels (Fig.~S1~c) cut out of a \SI{500}{\micro\meter} thick sheet of fluorinated ethylene propylene (FEP). To form a closed chamber, the sheet is sandwiched between two transparent sapphire plates of size \SI{22}{\milli\meter}\,$\times$\,\SI{60}{\milli\meter} with a thickness of \SI{1.0}{\milli\meter} (top) and \SI{0.5}{\milli\meter} (bottom) respectively. The three layers are pressed together by a steel frame screwed on an aluminum base to tightly seal the chamber. Hydrodynamic access to the chamber is provided by a total of four holes in the thin bottom sapphire that are connected to polytetrafluoroethylene (PTFE) tubings (ID\,\SI{18}{\micro\meter}) via standard low-pressure fittings, ferrules and corresponding threads in the aluminum base.

\subsection{Microscopy and Visualisation}
For visualisation of the chamber's interior the setup is placed on the stage of a microscope (Axio Zoom.V16, Carl Zeiss, Germany) using a three dimensional printed support frame. For brightfield imaging the chamber is illuminated by the microscopes built-in white light source from below, whereas for fluorescence imaging the sample is excited from above by the white light from a light engine (SOLA 80-10247, Lumencor, USA) coupled with a Zeiss 38 He filter to generate a \SI[parse-numbers = false]{470/40}{\nano\meter} light through the objective lens (0.5x). Images are recorded using a complementary metal–oxide–semiconductor (CMOS) camera (ORCA-Flash 4.0, Hamamatsu, Japan) or a mirrorless photographic camera (Lumix DC-G100K, Panasonic, Japan) for colour imaging mounted to the microscope with a C-mount adapter.

\subsection{Microfluidics}
The chamber is pre-filled manually with the ocean analog, whereas injection of the alkaline fluid is controlled by a pulsation-free, continuous-flow syringe pump (Nemesys S, Cetoni, Germany) operated with glass syringes (\SI{1}{\milli\liter}) at varying flow rates of \SI[parse-numbers = false]{1.0-15.0}{\milli\liter\per\hour}.

The FEP cutout (Fig.~S1~c) features channels of \SI{1.7}{\milli\meter} width that connect the rectangular cell to the holes in the sapphire. In preparation of the experiment, the chamber is pre-filled with ocean analog solution through the top outlet channels while keeping the inlet channels closed to avoid precipitation happening in the channels. To avoid bubbles in the chamber the alkaline solution is slowly pumped into the inlet channel (\SI{2}{\milli\liter\per\hour}) while simultaneously retrieving the air through the drainage channel at the same rate. All motion is stopped when the fluid front is \SI{2}{\milli\meter} from entering the chamber, the drainage channel is closed and the experiment is started at the desired flow rate.

\subsection{Sample Preparation}
The acidic ocean analog is prepared by dissolving iron(II)chloride tetrahydrate (\ce{Fe(II)Cl2 * 4H2O}) (Sigma-Aldrich, USA) in nuclease-free water (Ambion, Invitrogen, ThermoFisher, USA) to reach concentrations of either \SI{75}{\milli\molar} or \SI{200}{\milli\molar}. To keep oxidation to a minimum, the iron(II)chloride tetrahydrate is stored under inert gas (argon) and all flasks and vials are pre-flushed with argon gas. As alkaline fluid a commercial volumetric standard solution of sodium hydroxide (\ce{NaOH}) (Carl Roth, Germany) is either used as purchased (1M) or diluted with nuclease-free water to reach the desired concentration. Depending on the experiment the fluids are used directly, mixed in a ratio of 8:1 with a universal pH dye (KS90-UpHI, Lovibond, Tintometer, Germany) or mixed 50:1 with a solution of BCECF (2',7'-Bis-(2-Carboxyethyl)-5-(and-6)-Carboxyfluorescein) (Invitrogen, ThermoFisher, USA) for fluorescent imaging. For visualisation of fluid flow and particle accumulation the inflowing sodium hydroxide solution was mixed in a ratio of 18:1 with a 2.5\% aqueous suspension of \SI{10}{\micro\meter} fluorescent (bright blue) carboxylate polystyrene microspheres (Fluoresbrite, Polysciences, USA).

\section{Discussion}\label{sec12}
In this work, we showed that dominantly three different precipitation morphologies emerge in a quasi-2-dimensional alkaline vent model which have different abilities to facilitate the emergence of pH gradients and accumulation of dispersed particles. In our studies, we identified flow-rate and concentration of hydroxide-ions (i.e.~concentration of NaOH) in the inflowing fluid to be the critical parameters in structure formation and therefore focused on their variation in our experiments. Our data suggest that ``precipitation fingers", obtained at medium flow-rates and medium concentrations of alkaline fluid within our boundaries, yield the steadiest gradients in terms of steepness and stability. Crucial for the formation of pH gradients is the subdivision of the flow into narrower ``fingers" across the width of the chamber, that form strong mineral precipitates with the acidic solution in between. After reaching the morphological steady-state pH gradients are maintained by the continued inflow of hydroxide ions through all of the ``fingers". Although the ``layer/rings" structures exhibit strong precipitation barriers as well, they fail to generate gradients, since all of the ocean fluid is pushed out of the chamber by propagation of the precipitation front spanning across the whole chamber width (no acidic fluid enclosed). After some time a main flow is established as a short-cut through the morphologies, leading to fading fluorescence of the pH dependent dye in the adjacent areas. This probably forms weak pH gradients between main flow and outer areas that are, however, not separated by a barrier. The lack of barriers seems also the problem for the ``flow fields" as they hardly show any stable precipitate and therefore lack the ability to maintain gradients as the fluids just intermix through diffusion.

We observe that a change in the experimental parameters essentially corresponds to a shift of the morphology's formation timescale. This seems plausible, since precipitation happens more rapidly at higher concentrations and grows faster at higher flow-rates. Therefore, we found that ``layers/rings" form predominately within the first \SI{20}{\second}, then converge to the ``precipitation fingers" between roughly 30 and \SI{80}{\second} and finally fade into the diffusive fields after around \SI{100}{\second} as the local flow-velocity ceases and diffusive movement dominates.  In experiments exhibiting dominantly ``precipitation fingers" or flow fields, we can therefore observe small sections of the other two morphologies at the respective timescale. In order to promote the formation of gradients, ``precipitation fingers" should be extended as far as possible, which works best by ensuring stable precipitates using medium to high \ce{OH-} concentration and medium flow-rates.

In the experiments with fluorescent beads mixed into the inflowing solution, accumulation could only be deduced in the parallel ``precipitation finger" structures and the ``layers/rings", which can be seen by the increasing fluorescence intensity of the agglomerates. While some of the beads are already trapped during the mineral precipitation process, more are accumulated as long as the inflow continues. This could be due to diffusiophoretic effects \cite{Moller2017} or the advective flows through the resulting structures. However, in the ``layers/rings" the beads only seem to increase in the grooves and are thus not subject to either gradients or the continued inflow, after the shortcut-flow establishes.

In the field of chemobrionics, confined quasi-2-dimensional flow-setups (Hele-Shaw flow cells) are commonly used to study precipitation reactions and the resulting pattern formation. The experiments mostly use radial injection of the fluid \cite{Schuszter2016, Rocha2022, Wang2020, Balog2020, Bere2021, Comolli2021} rather than injection from one side as in our setting. Even though we used different injection and chemical composition, also e.g.~Rocha et al.~\cite{Rocha2022} showed the formation of different morphologies over the time of the experiment. However, none of the patterns identified in the referenced literature coincides exactly with our dominant morphologies, which would make our setting also interesting for modelling of the pattern formation process.

Modern alkaline vents systems, such as Lost City, can be seen as an example to our model by structure, however, not by chemical composition. We selected the fluids in order to resemble the crucial conditions that allow the precipitation of minerals upon contact and are assumed to have been prevalent on an early Earth. Although we did not aim to recreate the formation of any specific minerals, it is likely that we create white and green rust as shown in previous work under comparable conditions \cite{Helmbrecht2022}. Even though Lost City is considered a low-temperature alkaline vent system, the vented fluids still have temperatures of \SI[parse-numbers = false]{40-75}{\celsius} \cite{Kelley2001, Kelley2005}. Our results presented here were mainly focused on experiments carried out at room temperature, but as the supplementary data shows (see Fig.~S3), the injection of heated fluids seems not to have any significant influence on the resulting morphologies.

Our setup allows to study the alkaline vent formation process in quasi-2-dimensions, thereby providing full optical access for fluorescence and bright field visualisation and allowing for the precise control of chamber thickness and inflow-rate. All previous attempts to recreate the alkaline vent formation process in the lab were focused on a 3D setup \cite{Herschy2014, Barge2015}. The 2D representation serves as a model for the precipitation process in 3D and only requires sample volumes lower than \SI{100}{\micro\liter}. As the alkaline vent precipitates provide highly reactive mineral surfaces \cite{Barge2019, Helmbrecht2022} this would also provide a promising setting to study the effect of minerals and gradients on origins of life related chemical reactions in low volumes. Apart from that, the confined geometry could probably not only be seen as a model, but also as a potential geological scenario, considering the possibility that the vent precipitation process could already start in the cracks of the crust before the fluids are exhaled into the bulk ocean waters.

From an origins of life perspective, the quasi-2-dimensional scenario could satisfy several requirements, that are thought to be crucial for the emergence of living systems. Essentially, all life we know today depends on non-equilibria and, thus, requires gradients of ions and pH that are separated by a semi-permeable membrane \cite{Moller2017}. In the alkaline hydrothermal vent theory it is therefore considered that geo-chemical gradients could pose an early precursor to the topology of living systems, given their structural similarities to autotrophic cells \cite{Moller2017, Sojo2016}. This is also clearly reflected in our ``precipitation finger" morphology, with the alkaline fluid streaming inside the finger structures which are separated by a thin mineral barrier from the more acidic ocean solution in between the streams. For biochemical reactions, this setting could not only be beneficial because of the ionic gradient as a driving force, but also due to the reactive mineral surface with their potential catalytic abilities \cite{Moller2017, Sojo2016, Sleep2004}.
In order for chemical reactions to happen and to start molecular evolution, it requires high concentration of molecules and means to localise reagents and products to avoid dilution in an extensive ocean, an issue usually referred to as the `concentration problem' \cite{deDuve1991}. Additionally to the trapping of dissolved species in the physical structure of the morphologies (e.g.~formed pockets), the ion gradients could facilitate a directed colloid transport called diffusiophoresis \cite{Moller2017} to even further boost the accumulation of macromolecules. This would enhance the local concentration of dissolved molecules, which in turn could be localised by trapping or interaction with the minerals \cite{Helmbrecht2022} and eventually participate in reactions catalysed at the mineral surfaces.

\section{Conclusion}\label{sec13}

We investigated the precipitation of hydrothermal vent morphologies in a quasi-2-dimensional setting. We found that finger-like precipitates at medium flow-rate and \ce{OH-} concentration allow for the generation and maintenance of pH gradients and facilitate the accumulation of dispersed particles at the same time. Considering the possibility that alkaline vent precipitation could already happen within the fractures of the seafloor, the confined geometry could even be a promising mimic of geological scenario.

Our microfluidic setup allows us a spatio-temporal visualisation of the precipitation process and the emerging gradients with universal and fluorescent pH dyes. This could open the door for laboratory experiments in low volume to study the impact of alkaline vent environments on accumulation of organic molecules, catalytic reactions and processes like polymerisation or replication of bio-polymers.

\backmatter

\bmhead{Supplementary information}
Supplementary Figures 1-3 and supplementary Movies 1-9.

\bmhead{Acknowledgments}
Financial support is greatfully acknowledged from the European Research Council (ERC) Evotrap, grant no. 787356 (D.B.),the Deutsche Forschungsgemeinschaft (DFG, German Research Foundation) – Project-ID 364653263 – TRR 235 (CRC 235) (D.B.), and the Max Planck Society (K.A.).

We want to thank Lea S. Gigou for assistance in preliminary Experiments and Almuth Schmid and Christina F. Dirscherl for extensive comments on figures and manuscript.

\section*{Declarations}

\bmhead{Funding}
Financial support was provided by the European Research Council (ERC) Evotrap, grant no. 787356 (D.B.), the Deutsche Forschungsgemeinschaft (DFG, German Research Foundation) – Project-ID 364653263 – TRR 235 (CRC 235) (D.B.), and the Max Planck Society (K.A.).
\bmhead{Conflict of interest}
The authors declare no competing financial interest.
\bmhead{Ethics approval}
Not applicable.
\bmhead{Consent to participate}
All authors consent to their contribution.
\bmhead{Consent for publication}
All authors consent to publication of the final manuscript.
\bmhead{Availability of data and materials}
The collected experimental data, are available in the mediaTUM repository under doiXXX.
\bmhead{Code availability}
Not applicable.
\bmhead{Authors' contributions}
MW, DB and KA designed the experiments. MW, CD and SC performed all experiments. VH and WDO developed the sample composition and gave major input on geological interpretation. MW, SC, VH, WDO and KA analysed the data. KA and DB supervised the project. MW wrote the manuscript and all authors commented on the manuscript and contributed to the writing process.

\noindent

\bigskip


\end{document}